\documentclass[aip,jcp,reprint]{revtex4-1}
\usepackage{chemformula} 
\usepackage[T1]{fontenc} 
 \usepackage{float}


\begin{document}	
	\author{Amin Bakhshandeh}
	\email{amin.bakhshandeh@ufrgs.br}
	\affiliation{Instituto de F\'isica, Universidade Federal do Rio Grande do Sul, Caixa Postal 15051, CEP 91501-970, Porto Alegre, RS, Brazil.}
	
	\author{Derek Frydel}
		\email{derek.frydel@usm.cl}
	\affiliation{Department of Chemistry, Federico Santa Maria Technical University, Campus San Joaquin,7820275,  Santiago,Chile}

	\author{Yan Levin}
	\email{levin@if.ufrgs.br}
	\affiliation{Instituto de F\'isica, Universidade Federal do Rio Grande do Sul, Caixa Postal 15051, CEP 91501-970, Porto Alegre, RS, Brazil.}
	
	
	\title{Reactive Monte Carlo Simulations for Charge Regulation of Colloidal Particles}

	\begin{abstract}
		We use a reactive Monte Carlo simulation method and primitive model of electrolyte to study acid-base equilibrium that controls charge regulation in colloidal systems. The simulations are performed in a semi-grand canonical ensemble  in which colloidal suspension is in contact with a reservoir of salt and strong acid. The interior of colloidal particles  is modeled as a low dielectric medium, different from the surrounding water.  The effective colloidal charge is calculated for different number of surface acidic groups, pH, salt concentrations, and types of electrolyte.  In the case of potassium chloride the titration curves are compared with the  the experimental measurements  obtained using potentiometric titration.  A good agreement is found between simulations and experiments.  In the case of lithium chloride specific ionic adsorption is taken into account through partial dehydration of lithium ion.  	
	\end{abstract}
	\maketitle
	
	\newpage
	\section{Introduction}
	The interplay between electrostatic interactions~\cite{monica1,monica2,D1SM00232E,javidpour2019role,Christoph2020,Krishnan3,Krishnan1,Krishnan2,Krishnan4,Ion-Containing,Gao22030,avni2020critical,avni2019charge,ADAR2017198,SHEN200592,Tsao2000,L_wen_2005,Messina_l,MESSINA2002282,Borkovec2001,narambuena2021non} and acid-base equilibrium is of paramount importance in  Chemistry and Biology~\cite{worms2006b,booth1985regulation,antelo2005effects,brigante2016,hammes2002key,teixeira2010fast}.   It often defines the boundary between life and death.   The protein functionality can change due to modification of surrounding pH,  affecting  enzymatic activity  and the resulting metabolic processes~\cite{Kurkdjian}.  The  transport  of ions through the  cellular membranes ~\cite{accardi2004secondary} is strongly affected by pH.  The same is true for the  binding of heavy metals to bacterial membrane~\cite{beveridge1989role,van1997determination,heinrich2007acid}.
	In biological systems pH, which depends on the activity of proton inside the solution,  can be significantly modified by the presence of electrolyte, which at physiological concentration is around  $150$ mM~\cite{pressman1965induced,Lund2013},  affecting the  
	solubility and stability  of proteins~\cite{D0M_kosovan,lunkad2021role,perutz1978electrostatic,svensson1990electrostatic,woodward1991potentials,lund2005charge,dickinson2007food,da2009polyelectrolyte,teixeira2010fast,szuttor2021modeling,behrens2001charge,Lund2013,Mikaellund,lund2005charge}.  In colloidal science both pH and electrolyte concentration regulate colloidal charge and interaction between the particles~\cite{andelman2006,podgornik2018,Tsao2000,avni2019charge}. {\color{black}The Monte Carlo (MC) titration method which uses
		explicit ions has been successful in explaining of ionization equilibrium for complex systems such as protein
		complexation and charge regulation~\cite{teixeira2010fast,da2009polyelectrolyte,jonsson1996titrating,kesvatera1999ionization,andre2004role,labbez2009ion}.
		In these methods, explicit ions are added and removed in constant pH to the system.  }


	In the present paper we will use reactive Monte Carlo simulation  to efficiently explore the effective charge of colloidal particles inside an electrolyte solution of a given pH.  Differently from constant pH simulations~\cite{svensson1990electrostatic,woodward1991potentials}, we  treat all ions,  including hydronium, explicitly in the semi-grand canonical ensemble.   The advantage of this approach is that  it automatically enforces uniform electrochemical potential of all ion inside the simulation cell~\cite{madurga2011semi,nishio1996monte,nishio1994monte,johnson1994reactive,valleau1980primitive,lan}. 	
	In this respect our approach is similar to other recently introduced grand canonical simulation methods to account for charge regulation ~\cite{Fabian,Luijten,Morten,lan,madurga2009ion,madurga2011semi}.
\textcolor{black}{However, the exclusive treatment of all ions makes the method general and so it can be used to treat other 
surface groups, not just those involving protonation.  For example, in this work, we consider adsorption of \ch{Li+}.   Similarly, the method could be used 
to treat basic surface groups.  This leads to the consistency in the method as there is no need to switch from method to method
to simulate different surface groups.}

	
	
	This article is organized as follows.  In Sec. (2) we introduce and describe the system.  
	In Sec. (3) we discuss the simulation method.  In Sec. (4) we compare the 
	simulation method with the experimental results.  In Sec. (5) we consider the scenario in which a cation \ch{K+} is substituted 
	by \ch{Li+}, which can specifically adsorb to carboxylate.  The work is concluded in Sec. (6).

	\section {Model and system details}
	\label{sec:sec1}
	
\begin{figure} 
		{\includegraphics[width=4.8cm]{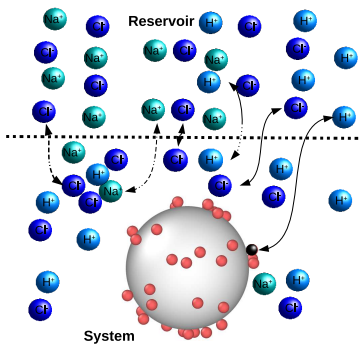} }
		\caption{Schematic representation of reactive MC moves. The deprotonated  acidic groups \ch{A-} \textcolor{black}{randomly} 
		distributed over the colloidal 
			surface are represented by the red spheres and protonated  \ch{HA} groups are represented by the black spheres.  
			The protonation of a titration site, \ch{A-} + \ch{H+} $\rightarrow$ \ch{HA}, changes its  charge from  $-1q \rightarrow 0$, 
			at the same time \ch{Cl-} is added to the cell with grand canonical probability, to preserve the overall charge neutrality.  
			Deprotonation reaction,  \ch{HA} $\rightarrow$ \ch{A-} + \ch{H+}, results in the change $0 \rightarrow -1q$ of the titration 
			site,  and a simultaneous removal of \ch{Cl-} from the simulation cell. }
		\label{f1}
	\end{figure}

	The system consists of a single colloidal particle of radius $a$ fixed at the center of a spherical Wigner-Seitz (WS) cell 
	whose radius $R$ determines the volume fraction of the colloidal suspension of a corresponding experimental system.  
	The cell is allowed to exchange ions with a reservoir containing strong monovalent salt and acid, 
	at concentrations $c_s$ and $c_a$, respectively, both of which fully dissociate in an aqueous solution, see Fig. \ref{f1}.    
	All ions have charge $\pm q$, where $q$ is the proton charge.  The dielectric constant inside the colloidal sphere, 
	representing a polystyrene latex particle, is $\epsilon_c \approx 2.5$, while the dielectric constant of the surrounding 
	water is $\epsilon_w \approx 80$.   One advantage of working with the WS formalism is that contact theorem allows us to 
	calculate the osmotic pressure inside the colloidal suspension directly 
	using ionic densities at contact with the cell boundary~\cite{wenncell}.
	
	The colloidal surface contains acidic 
	groups of weak acid that only partially dissociate upon contact with an aqueous solution: the process is 
	governed by the reaction \ch{HA <=> H+ + A-} and the corresponding dissociation constant $K_a$.  These surface 
	groups are modeled as small spheres of radius $r_0=2$\AA\ \textcolor{black}{randomly} distributed over the colloidal surface.  
	(\textcolor{black}{The results are insensitive to the mode of distribution or the grade of randomness}).  
	The dielectric constant 
	of small spheres is the same as that of water, so that dielectric discontinuity occurs only at the colloidal boundary.  
	The choice to keep the smaller spheres lying on the colloid surface rather than embedded into it has a physical motivation.  
	Charged surface groups prefer being displaced from the region of low dielectric constant.  
	The acidic surface groups are also referred to as titration sites.  In its deprotonated state a titration site has charge 
	$-q$, and in a protonated state it has charge $0$, see Fig. \ref{f1}.  The charge of titration site is located at its center.  
	
	The region outside the colloidal particle is occupied by the ions of a fully dissociated strong salt and acid, 
	\ch{KCl -> K+ + Cl-} and \ch{HCl -> H+ + Cl-}.  All ions are represented by hard spheres of radius 
	$3.3$~\AA~\cite{nightingale1959}, corresponding to fully hydrated ions.  In particular, a proton 
	$\ch{H+}$,  does not exist separately, but is a hydronium ion, $\ch{H3O+}$.  
	

	The dielectric contrast between the colloidal core and the surrounding water results in induced surface charge density at the interface.  
	The surface charge distribution depends on the location of ions inside the suspension.  Therefore, the interaction between any two ions 
	near a colloidal particle will not be given simply by the Coulomb potential, but will also depend on their distance from the colloidal surface.  The 
	Green function for the effective interaction between two ions inside the WS cell is accurately 
	approximated by~\cite{bkh2011,dos2011,BKH2018}
	
	\begin{equation} 
		G({\bf r},{\bf r'})=\frac{1}{\epsilon_w |{\bf r}-{\bf r'}|}+\frac{ \gamma a}{\epsilon_w r'|{\bf r}-\frac{a^2}{r'^2}{\bf r'}|}+\gamma \psi_c({\bf r},{\bf r'}),
		\label{1}
	\end{equation}	
	with
	\begin{equation}
		\psi_c({\bf r},{\bf r'})=\dfrac{1}{\epsilon_w a}
		\ln \left[\frac{r r' -{\bf r} \cdot {\bf r'}}{a^2-{\bf r} \cdot {\bf r'}+\sqrt{a^4-2 a^2 ({\bf r} \cdot {\bf r'})+r^2 r'^2}}\right] \ ,	
	\end{equation}
	where $\gamma=\left(\epsilon_w-\epsilon_c\right)/\left(\epsilon_w+\epsilon_c\right)$.   The first term of Eq.~\ref{1} is due to 
	the direct Coulomb potential produced by an ion located at  position  ${\bf r'}$ from the center of colloidal particle, while  
	other two terms result from the induced surface charge~\cite{norris1995charge,lindell1993image}.  Note that the Green 
	function is invariant under the exchange of the source and the observation points ${\bf r} \leftrightarrow {\bf r'}$.   Eq.~\ref{1} 
	is exact in the limit  $ \epsilon_c/\epsilon_w \rightarrow  0$, $\gamma=1$.  Furthermore, it was shown that it remains very 
	accurate for  $\epsilon_c/\epsilon_w \ll 1$~\cite{dos2011}, which is appropriate for the present system with  $\epsilon_c/\epsilon_w =2.5/80$. 
	
	The expression in Eq. (\ref{1}) is derived specifically for a spherical colloidal particle.  
	To account for dielectric polarization in 
	more complex geometries, one would need to resort to numerical boundary-element methods, such as the Iterative Dielectric Solver (IDS) \cite{Luijten}, which is much more computationally intensive.

	The electrostatic energy of the system is: 
	\begin{equation} 
		\begin{aligned}
			U= \gamma\sum_{i=1}^{N}\left[\frac{q^2 a}{2 \epsilon_w \left(r_i^2 -a^2\right)} +\frac{q	\psi_{self}({\bf r_i})}{2}\right] +\\
			\sum_{i=1}^{N-1}\sum_{j=i+1}^{N} q _i q_j	G({\bf r_i},{\bf r_j}),
		\end{aligned}
		\label{2}
	\end{equation}
	where $N$ is the total number of particles inside the cell, including the charged surface sites.  
	The term in the square brackets is due to self-interaction of an ion with its own induced surface charge and vanishes 
	for $\gamma=0$, where $\psi_{self}$  is given by
	\begin{equation}
		\begin{split}
			\psi_{self}({\bf r_i}) = \frac{ q  }{\epsilon_w a  } \ln{(1-\frac{a^2}{r_i^2})}.
		\end{split}
		\label{Eq7}
	\end{equation}
	Hard-sphere interactions between ions is not included in (\ref{2}).   Configurations with 
	overlapping hard-spheres are not allowed.  Since  the hydrated size of all ions is the same, \ch{H3O+} and \ch{K+} are 
	identical.  The difference between the two ions is that \ch{H3O+} can react with acidic groups.  When 
	reacting,  it transfers its charge to them, while the remaining neutral hard-sphere disappears.  
	This is physically justifiable since hard-sphere in our model represents a hydrated structure, which vanishes
	upon the proton transfer.

	\section{Simulation Method}
	\label{sec:sec2}
	
	The simulations are performed in the semi-grand canonical ensemble,  in which the system is in contact with an infinite reservoir of strong acid and salt. Since the WS cell represents an infinite colloidal suspension at finite concentration, we may think of it as being separated from the reservoir by a semipermeable membrane which allows for a free passage of ions, but not of colloidal particles.   The  diffusion of ions will result in an electric field across the membrane, establishing a potential difference between the system and the reservoir.  This is known as the Donnan potential, $\varphi_D$.  When an ion  moves from the reservoir into the system  it will, therefore,  gain additional energy $q_i \varphi_D$.   
	
	Our reactive Monte Carlo involves two types of movements.  The standard grand canonical insertion/deletion of ions between the reservoir and the simulation cell;  and protonation and deprotonation moves for the surface sites.  The grand canonical acceptance probabilities  for insertion or deletion of an ion of type $i$ are given by $ACC = \min\left(1,\phi_{add/rem}\right) $, where $\phi_{add/rem}$ is
	\begin{equation}\label{eqd1}
		\begin{split}
			\phi_{add} =  \frac{V c_i}{N_i +1}\exp\left[- \beta\left( \Delta E_{ele}- \mu_{ex}+ q_i\varphi_D  \right) \right],\\
			\phi_{rem} =  \frac{ N_i}{V c_i}\exp\left[- \beta\left( \Delta E_{ele}+ \mu_{ex}- q_i \varphi_D  \right) \right].
		\end{split}
	\end{equation}
	Here $\Delta E_{ele}$ is the change in electrostatic energy upon insertion or deletion of an ion,  $c_i$ is the concentration of ion of type $i$ in the reservoir,  $\mu_{ex}$ is the excess chemical potential of ions in the reservoir, and $V$ is the free volume accessible to the ions.   Note that since all ions are assumed to have the same radius and are all monovalent, the excess chemical potential of all ions is the same.  The Donnan potential   $\varphi_D$  has to be adjusted each few MC steps to keep charge neutrality inside the system~\cite{Panagdonnan}.   This makes simulations slow.  A simple solution to this problem is to perform insertion or deletion of cations-anion pairs,  so that the charge neutrality of the cell is always preserved.  For example, the  probability of  insertion/deletion  of \ch{KCl} pair is the product $\phi_{add/rem}^{\ch{K+}}\phi_{add/rem}^{\ch{Cl-}}$,  so that the Donnan potential cancels out.

	In the grand canonical ensemble concentrations are controlled indirectly via chemical potentials.  
	One way to establish the correspondence between $\mu_{ex}$ and the ionic concentrations is to run simulations 
	for a bulk solution at a given concentration and then compute $\mu_{ex}$ using, for example, the Widom insertion 
	procedure.  Alternatively one can perform a grand canonical MC with fixed chemical potentials and observe the corresponding concentration of ions inside the simulation box.  
	We note, however, that our ionic solution is a restricted primitive model.  We can, therefore, take advantage 
	of this fact and use an accurate analytical approximation for the excess chemical potential given by 
	$\mu_{ex} = \mu_{CS} +\mu_{MSA}$, where 
	$\beta\mu_{CS} = \frac{8\eta-9 \eta^2+3\eta^3}{\left(1-\eta\right)^3}$ 	is the Carnahan-Starling expression for the excluded volume 
	contribution~\cite{carnahan1969equation,carnahan1970thermodynamic,adams1974chemical,MACIEL2018}, 
	where the volume fraction is $\eta=\frac{\pi d^3}{3} c_t$,  $d$ is the ionic diameter, and  $c_t=c_s+c_a$ is the total concentration of salt and acid.
	The electrostatic contribution to the excess chemical potential can be accurately calculated using the mean-spherical approximation (MSA)~\cite{ho1988mean,ho2003interfacial,levin1996criticality,waisman1972mean,waisman1972mean2,blum1975mean}, 
	\begin{equation}
		\label{eq4}
		\beta \mu_{MSA} = \frac{\lambda_B\left( \sqrt{1+2 \kappa d}-\kappa d -1\right)} {d^2\kappa},\\
	\end{equation}
	where   $\lambda_B=q^2/\epsilon_w k_B T$ is the Bjerrum length, and  $\kappa= \sqrt{8 \pi \lambda_B c_t}$ is the inverse Debye length.
	
\begin{figure}[H]
		\centering
		\includegraphics[width=0.6\linewidth]{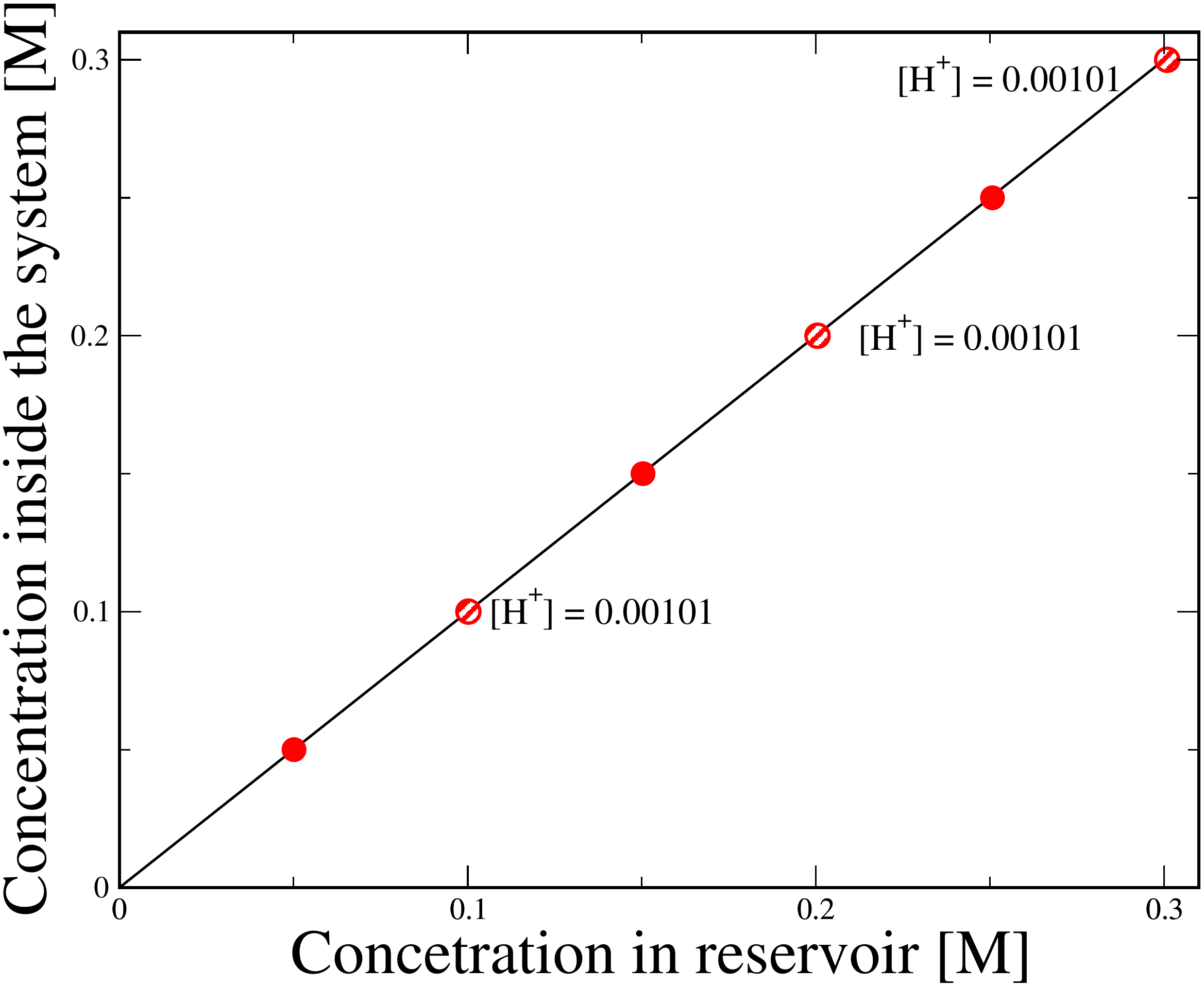}
		\caption{Salt concentration inside the system as a function of the reservoir concentration.  The reservoir also contains strong acid at concentration $c_a=0.001$M.  The straight line indicates the theoretical expectation that the concentrations of ions inside the system should be exactly the same as the concentration of ions in the reservoir.  The concentration of hydronium ion inside the system is indicated next to the symbols. The  excellent agreement between concentrations inside the system and in the reservoir indicates accuracy of our expression for $\mu_{ex}$. 
		}
		\label{fig:saltmsa}
	\end{figure}

	
	To test this procedure, we choose concentrations of acid and salt from  which we  calculate $\mu_{ex}$
	using the analytical expressions above.  We then run a grand canonical MC 
	simulation for this value of $\mu_{ex}$, starting with an empty simulation cell.  
	If the approximate expression for $\mu_{ex}$ is accurate, once the simulation is converged, we should recover the concentration of acid and salt inside the simulation cell that we specified to calculate $\mu_{ex}$.  
	This is precisely what is observed in the simulations, see Fig. \ref{fig:saltmsa},  justifying the use of the analytical expression for $\mu_{ex}$.	
	
	To obtain the MC weights for protonation/deprotonation moves,  we first consider the free energy change for a reaction of a hydronium ion with an isolated weak acid group \ch{A-}, \ch{H3 O+ + A- <=> HA  + H2 O}.
	We can think of this process as a removal of a hydronium  ion from the reservoir, followed by a reaction with a \ch{A-}  group.  The internal partition function of the \ch{HA} molecule is  $K_{eq}/\Lambda^3_{\ch{H+}}$, where   $K_{eq}$ is the equilibrium constant and $\Lambda_{\ch{H+}}$ is the hydronium's de Broglie thermal wavelength.  The free energy change  of the system$+$reservoir resulting from the removal of a hydronium ion from the reservoir and bringing it into
	vicinity of \ch{A-}, where it will undergo a proton transfer reaction,   is  then $\beta \Delta F_p=-\ln(K_{eq}/\Lambda^3_{\ch{H+}})-\mu_{\ch{H+}}$, where $ \beta \mu_{\ch{H+}}=\ln(c_{\ch{H+}}\Lambda^3_{\ch{H+}})+ \beta \mu_{ex}$ is the total chemical potential of a hydronium ion inside the reservoir.  For the deprotonation reaction the free energy change is $\Delta F_d=-\Delta F_p$. 	The acceptance probabilities for protonation and deprotonation moves are, therefore,
	\begin{equation}\label{eqRM_1}
		\begin{split}
			\phi_p =  \exp\left[-\beta \left(\Delta E_{ele}  +  \Delta F_p    +q\varphi_D   \right)\right] ,\\
			\phi_d = \exp\left[-\beta \left(\Delta E_{ele} +\Delta F_d  -q\varphi_D     \right)\right],
		\end{split}
	\end{equation}
	respectively.  These simplify to
	\begin{equation}\label{eqRM}
		\begin{split}
			\phi_p =  c_{\ch{H+}}K_{eq}\exp\left[-\beta \left(\Delta E_{ele}  -  \mu_{ex}    +q\varphi_D   \right)\right],\\
			\phi_d = \frac{1  }{c_{\ch{H+}}K_{eq} }\exp\left[-\beta \left(\Delta E_{ele} +  \mu_{ex} -q\varphi_D     \right)\right] .
		\end{split}
	\end{equation}
	A Monte Carlo ``reaction" move consists of selecting a random 
	titration site,  followed by an attempt to change its ``state" from protonated to deprotonated and vice versa.  
	Note that the acceptance probabilities depend on the Donnan potential, which is {\it a priori}, unknown.  We can overcome this difficulty by again combining a protonation attempt with an insertion of \ch{Cl-}.  The probability of acceptance of a protonation-\ch{Cl-}-insertion move is then $\text{min}\{1,\phi_{p}\phi_{add}\}$.  In the product, the Donnan potential once again cancels out.  Similarly a deprotonation move is combined with a removal of \ch{Cl-}, so that the probability is $\text{min}\{1,\phi_{d} \phi_{rem}\}$.  The final acceptance probabilities for the pair moves are:
	\begin{equation}\label{eq7}
		\begin{split}
			\phi_{d+rem} = \frac{ N_{\ch{Cl-}}}{c_{\ch{H+}}K_{eq}V c_{\ch{Cl-}}}\exp\left[-\beta \left(\Delta E_{ele} +2\mu_{ex}      \right)\right] ,\\
			\phi_{p+add} = \frac{ c_{\ch{H+}}K_{eq} V c_{\ch{Cl-}}}{(N_{\ch{Cl-} } +1)}\exp\left[-\beta \left(\Delta E_{ele} -2\mu_{ex}        \right)\right]
		\end{split}
	\end{equation}
	where $N_{\ch{Cl-} }$ is the number of \ch{Cl-} ions inside the cell.  In Fig.~\ref{f1} we have schematically shown these movements.

	\section{Comparison with the experimental results}		
	\label{sec:sec3}

	We now apply the simulation method discussed above to study the effective charge of carboxyl latex particles~\cite{behrens2000charging},
	the surface sites of which undergo reaction \ch{--COOH <=> COO- + H+}.
	The simulated system consists of a spherical cell of  radius  $R=120$\AA, with 
	a spherical colloidal particle at the center, represented by a hard-sphere of 
	radius $a=60$\AA, with surface titration sites of $r_0=2$\AA, see Fig. \ref{f1}.   
	
	The mobile ions correspond to the dissociation of \ch{KCl} and \ch{HCl}.  We are interested in obtaining the effective charge of a colloidal particle as a function of pH and salt concentration in the reservoir.  Recall that pH$=-\log[a_{\ch{H+}}]$,
	where $a_{\ch{H+}}=c_{\ch{H+}} \exp[\beta \mu_{ex}(c_t)]$ is the activity of hydronium ions.  Furthermore,  the electrochemical potential of each ions inside the system and in the reservoir are the same.  Note that inside the  simulation cell electrochemical potential includes the Donnan potential.  This means that activity of hydronium ion is the same inside the cell and in the reservoir, therefore, pH inside the simulation cell and inside the reservoir are also the same. 
	
\begin{figure} 
		\includegraphics[width=.8\linewidth]{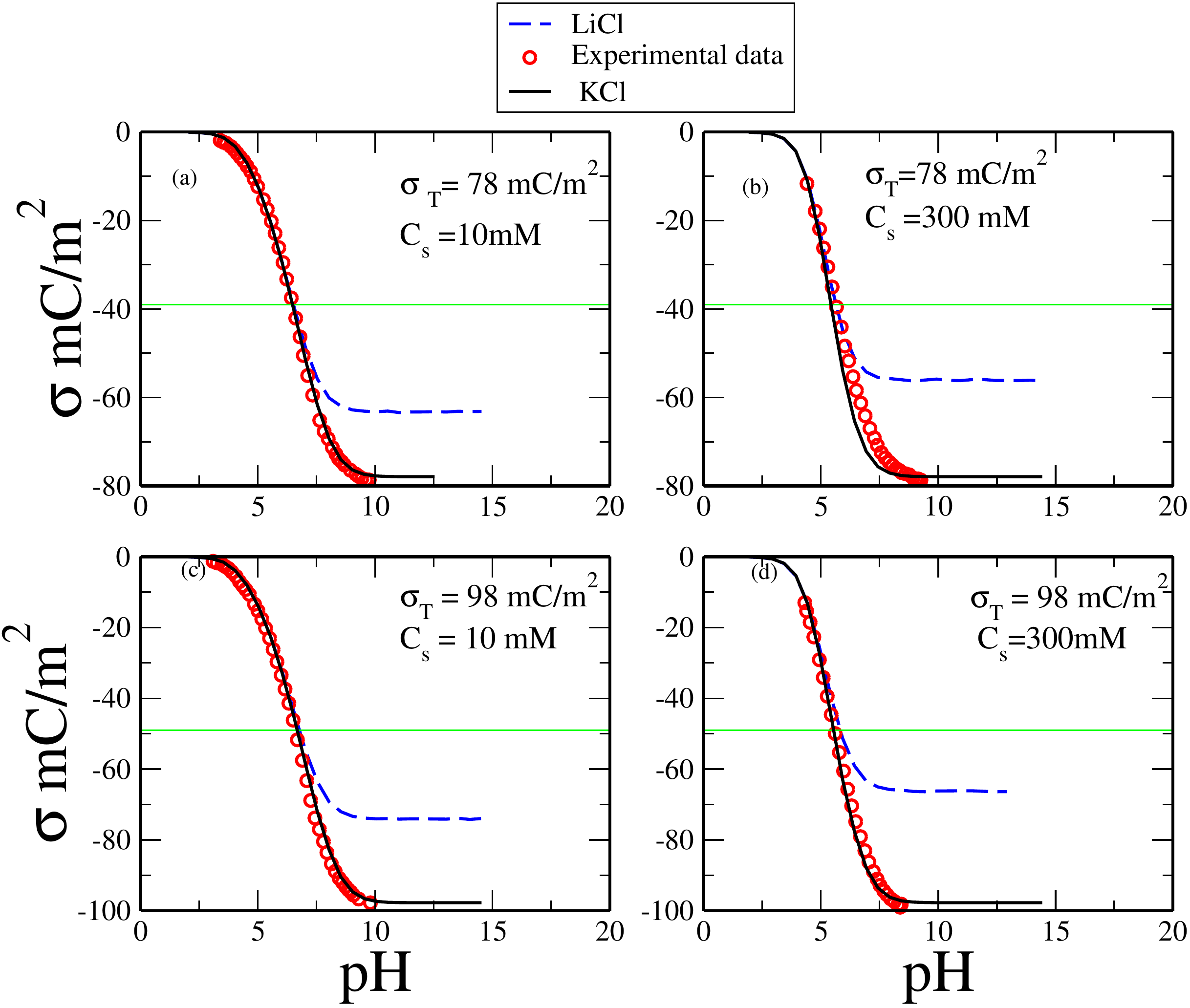}
		\caption{Effective charge obtained using the reactive MC simulations (curves) compared with the  experimental data (symbols). The solid black curve is the effective charge in the presence of \ch{KCl} and dashed green curve in the presence of \ch{LiCl}. The bare surface charge densities of colloidal particles are $78$ mC/m$^2$ panels (a) and (b); and $98$ mC/m$^2$ panels (c) and (d). Electrolyte concentration is indicated in each panel.  The surface equilibrium constant, p$K_a=5.02$ is obtained by fitting the experimental data in the panel (a). The {\it same} equilibrium constant is then used to calculate the effective charges for other salt concentrations and colloidal bare charge: panels (b),  (c) and (d). A very good agreement is observed between theory and experiment~\cite{behrens2000charging}.}
		\label{fig:titration}
	\end{figure}

	We consider two different carboxyl latex colloidal particles,  the effective charge of which has been obtained experimentally using potentiometric titration~\cite{behrens2000charging}.
	It is important to remember that the surface association constant $K_{eq}$ is different from the bulk association constant for the same acid~\cite{buffle1988complexation,cera_1995,Garc,bakhshandeh2019,bk2020,CR2021}.  This difference may be attributed to the distinct symmetry of the electronic wave function of a chemical group  when it is on the surface of colloidal particle and  when it is in the bulk.
	Our strategy, then, is to adjust $K_{eq}$  to fit the experimental titration curve for colloidal particles with the bare surface charge density $\sigma_{T} = 78$ mC/m$^2$  -- determined by the total number of surface carboxyl groups   \textcolor{black}{and amounting to about $~900$ titration sites distributed over the colloid}
	--  inside the electrolyte solution of c$_{\ch{KCl}}$ = 10 mM.  We will then use the {\it same} $K_{eq}$ to calculate the 
	theoretical titration curves for other salt concentrations and  different colloidal bare charge.    Recalling that acid dissociation constant is $K_a=1/K_{eq}$,  we obtain an excellent fit of the experimental titration curve using p$K_a=-\log[K_a]=\log[K_{eq}]=5.02$, see Fig. \ref{fig:titration}(a).  This value is actually very close to the p$K_a$ of bulk acetic acid. 
	We next use the same value of p$K_a$ to calculate the titration curve for a particle of the same bare surface charge inside 
	an electrolyte solution of $300$mM.  Only a reasonable  agreement between experiment  and theory is obtained for this case, Fig. \ref{fig:titration}(b),  with a significant deviation appearing at large pH.   Finally, we calculate the titration curves for colloidal particles of surface charge $\sigma_T=98$ mC/m$^2$, in solutions of $10$mM and $300$mM electrolyte, using the {\it same} surface 
	equilibrium constant as before.  A good agreement is found between simulations and experiments for both of these concentrations of electrolyte, see Fig. \ref{fig:titration} panels (c) and (d).   At the moment we do not have a clear understanding of what is the cause of  the deviation between theory and experiment observed for particles of surface charge  $\sigma_{T} = 78$ mC/m$^2$ inside $300$mM electrolyte solution.  The deviation is particularly surprising in view of the fact that for $\sigma_T=98$ mC/m$^2$ and $300$mM electrolyte we do have a good agreement between theory and experiment.  
	
\begin{figure} 
		\includegraphics[width=.6\linewidth]{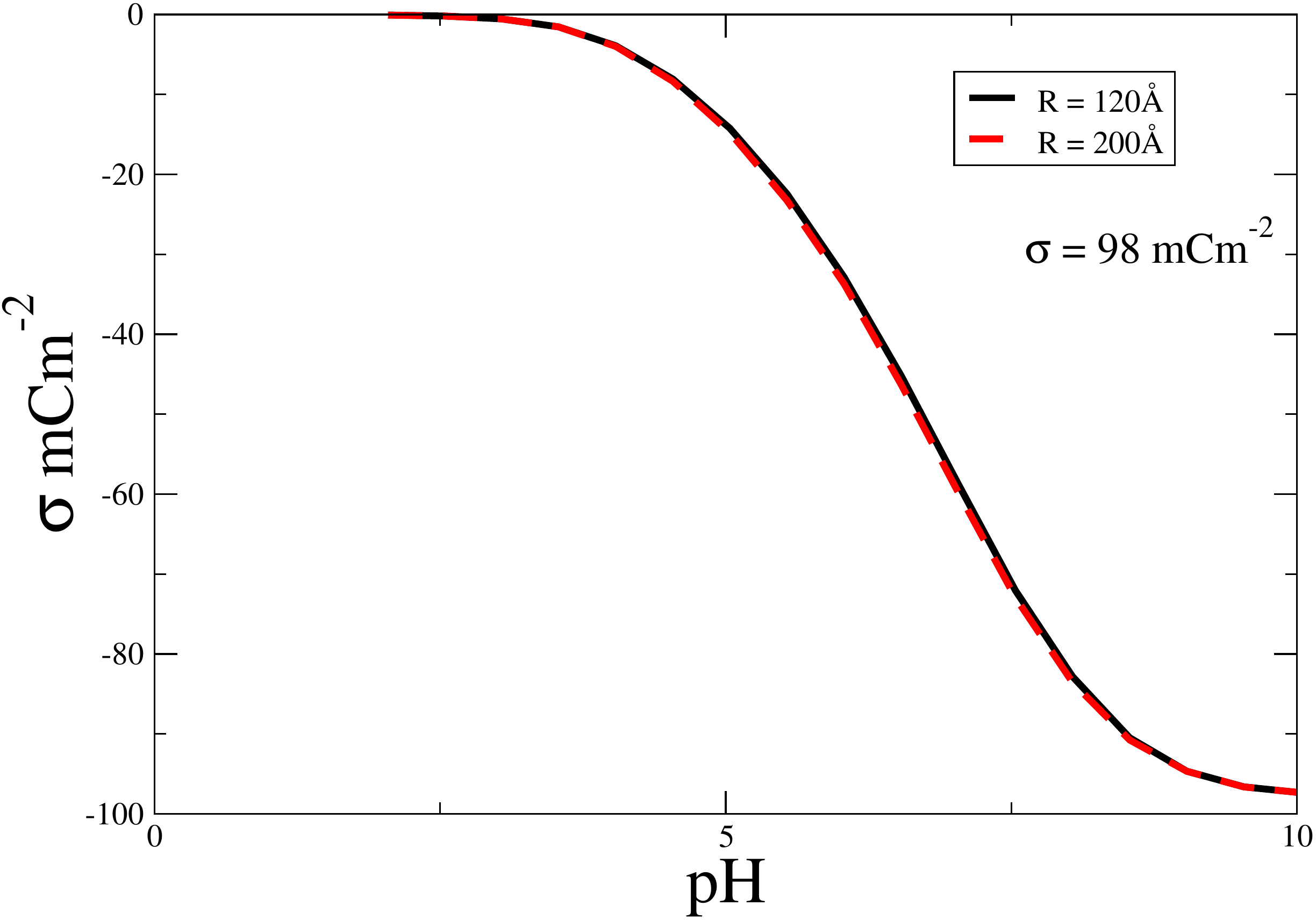}
		\caption{Titration curves for a particle with surface charge density $98$ mC/m$^2$ inside two different WS cells.}
		\label{fig:2R}
	\end{figure}

	We should note that in our simulation the size of colloidal particles is much smaller than in experiment.  We have checked, however, that this does not affect the titration curve, as long as the charge density of the particles is the same.  Similarly our calculations are performed for colloidal suspension at finite volume fraction, while experiments were done at infinite dilution.  Again the size of the WS cell does not affect the titration curve, as long as the cell is sufficiently large.  To demonstrate this in Fig. \ref{fig:2R} we present the titration curve for particles with bare surface charge density  $98$ mC/m$^2$ inside two different WS cells of radius $120$\AA\ and $200$\AA.  The two titration curves are practically indistinguishable.  This shows that the effective colloidal charge does not depend of the volume fraction, for sufficiently dilute suspensions.	

	\section{Specific ion adsorption}		\
	\label{sec:sec4}
	
	The specific  ion adsorption can be easily incorporated within the simulation formalism introduced above.  It is known that \ch{Li+}  specifically associate with \ch{COO-}~\cite{Uejio6809,sthoer2019}.  We are, therefore, interested to explore the effect of replacing \ch{KCl} with \ch{LiCl} salt.  The specific association of  \ch{Li+} with \ch{COO-} is a consequence of the law of matching water affinities (LMWA)~\cite{LMWA}, which suggests that both lithium and carboxylate can lose part of their hydration sheath resulting in strong electrostatic interaction between the two ions~\cite{biom5042435,winstein1961salt,record1976ion,lu2005separated,masnovi1985direct,yabe1992contact}.
	Based on the LMWA we will take the distance of closest approach between lithium and carboxylate contact pair to be $d=2.9$\AA, corresponding to the Latimer diameter of  \ch{Li+}  ion~\cite{diehl2012surface,latimer1993free}.  We can then say that any lithium ion that is within a distance $d=5.3$\AA\  --- the separation distance between a fully hydrated \ch{Li+} and the adsorption site  --- is associated with the carboxylate group, and will contribute to the renormalization of the effective charge.
	
\begin{figure}[H]
		\centering
		\includegraphics[width=0.6\linewidth]{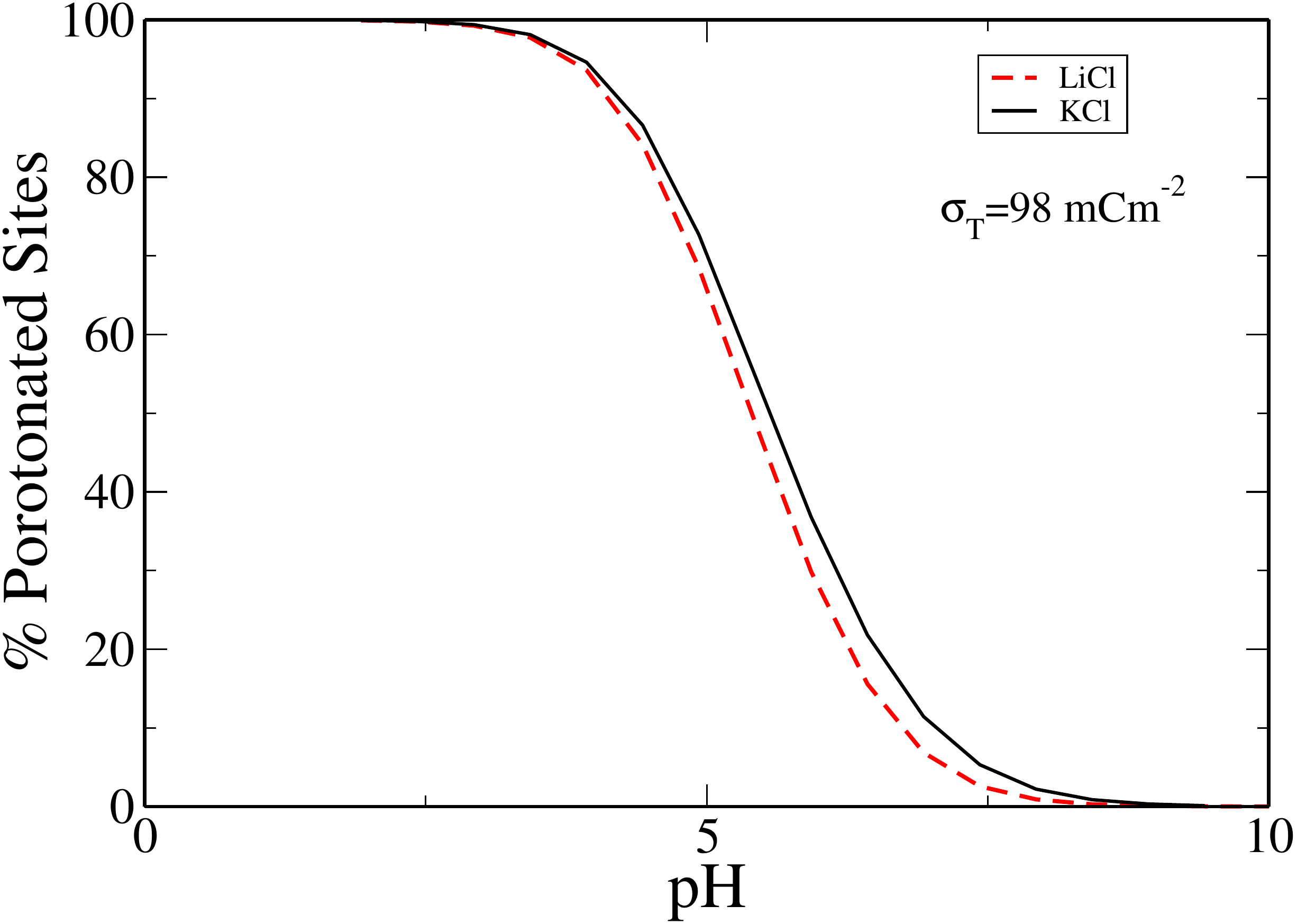}
		\caption{The fraction of protonated groups in the presence of $300$mM of either \ch{KCl} or \ch{LiCl}  salts.  Specific adsorption of \ch{Li+}  to carboxylate groups results in stronger deprotonation, in particular at large pH.}
		\label{f4}
	\end{figure}

	In  Fig.~\ref{fig:titration} (dashed green curves) we show the effect of  replacing \ch{K+}  with  \ch{Li+} on the effective colloidal charge.   Note the saturation of colloidal charge at large pH,  resulting from the association of \ch{Li+} ions with the carboxylate groups.  Specific adsorption of \ch{Li+} also affects the protonation of carboxylate, although the effect appears to be quite small, see Fig.~\ref{f4}.   We expect to see a more significant effect if \ch{Li+} is replaced by divalent \ch{Ca^++} ion, which is known to strongly interact with the carboxylate.  This will be explored in the future work.
	
	The degree of protonation for the same pH is lower in the case of \ch{LiCl} compared to \ch{KCl} salt, since \ch{Li+} ion effectively competes 
	with \ch{H+} for acidic sites.  Although deprotonation is larger for \ch{LiCl}, this does not imply that the effective colloidal surface charge is also larger.   Adsorbed \ch{Li+} ions neutralize sites  similarly to  \ch{H+} and, therefore, contributed to the effective charge of colloidal particles.  Thus, the effective surface
	charge of colloidal particles in the case of \ch{LiCl} solutions is lower in modulus compared to solutions of \ch{KCl}, for which no specific association takes place, see Fig.~\ref{fig:titration}.  Again we expect the effect of specific association to be even more important for suspensions containing  \ch{CaCl2} .  
	{\color{black} One quantity which normally use in Biochemistry and analytical chemistry is  Henderson–Hasselbalch (HH) equation which relates $	\text{pKa}^{HH}$ to $\text{pH}$ and the ration of active and titrated molecules of system, however when half of titration is done one can have:   
		\begin{equation}~\label{HH} 
		\text{pK}_a^{HH} =\text{pH}_{1/2}+\varphi \log10[\mathrm{e}],
		\end{equation}
Using the Eq.~\ref{HH} we compared the value of obtained from titration curve and HH equation which can be seen in Table below:
\begin{table}
\begin{tabular}{ccc}
	\hline
	$\sigma$ mCm$^{-2}$ ~~&~~ KCl mM & ~~~~$\text{pk}_a^{HH}$ \\
	\hline
	78 & 10 & 5.47 \\
	\hline
	78 & 300 & 5.25 \\
	\hline
	98 & 10 & 5.31 \\
	\hline
	98 & 300 & 5.32 \\
	\hline
\end{tabular}
\caption{\label{tab:table-name}The values obtaied from HH equation. The value was fitted in simulation $\text{pk}_a$ is 5.02.}
\end{table} 	
  }
	

	\section{Conclusions}	
	\label{sec:sec5}

	We have presented a simulation method which allows us to very accurately calculate the titration curves for suspensions of colloidal particles. 
	The results were  compared with the experimental measurements of the effective colloidal charge obtained using potentiometric titration.  A good agreement was found between simulations and experiments.   We have also shown how the specific ion interaction, responsible for the Hofmeister effect,  can be easily included in the simulation method.  
	The approach presented in this paper can be also applied to other scientifically and technologically important systems, such as proteins, polyelectrolyte gels,  polyampholytes, etc.  
	\section{Acknowledgments}
	This work was partially supported by the CNPq, CAPES, National Institute of Science and Technology Complex Fluids INCT-FCx. The authors are grateful to the Instituto de F\'isica e Matem\'atica, UFPel for the use of computer resources. D.F. acknowledges financial support from FONDECYT through grant number 1201192. 
	
	\bibliography{ref}
	
\end{document}